\documentclass[12pt]{article}
\usepackage{graphicx}

\begin{document}
\title{$K_{l4}$ decays}
\author{Bing An Li\\
Department of Physics and Astronomy, University of Kentucky\\
Lexington, KY 40506, USA}

\maketitle
 
\begin{abstract}
An effective theory of large $N_{C}$ QCD of mesons
has been used to study six $K_{l4}$ decay modes.
It has been found that the matrix	
elements of the axial-vector current dominate the $K_{l4}$ decays.
PCAC is satisfied. A relationship between three form factors of
axial-vector current has been found. Non-zero phase shifts are originated
in $\rho\rightarrow\pi\pi$.
The decay rates are calculated in the chiral limit.
In this study there is
no adjustable parameter.

\end{abstract}
\newpage
\section{Introduction}
There is rich physics in kaon decays. Study on
rare kaon decays are still active. The
theoretical study
of $K_{l4}$ decays has a long history[1,2].
 
In Ref.[3] we have proposed an effective theory of large $N_{C}$ QCD[4] of
mesons. In this theory the diagrams at the tree level are at the leading
order in large $N_{C}$ expansion and the loop diagrams of mesons are at higher
orders.
This theory is phenomenologically successful[5,6,7].
We have used this theory to study $K_{l3}$[3],
$K\rightarrow e\nu\gamma$[5,8], kaon form factors[7],
and $\pi$K scattering[6] in the chiral limit.
Theoretical results agree well with
data. In these studies VMD and PCAC are satisfied.
There are five parameters in this theory: three current quark masses,
a parameter related to the quark condensate, and a universal coupling
constant g which is determined to be 0.39 by fitting $\rho\rightarrow ee^+$.
All parameters have been fixed by previous studies.
 
In this paper we use this theory
of pseudoscalar, vector, and axial vector mesons[3] to study
$K^{-}\rightarrow\pi^{+}\pi^{-}l\nu, \pi^{0}\pi^{0}l\nu$, and
$K_{L}\rightarrow\pi^{\pm}\pi^{0}l^{\mp}\nu$.
In the study of $K_{l4}$ there is no
adjustable parameter.
 
The Lagrangian of the theory of Ref.
3Y is
\begin{eqnarray}
{\cal L}=\bar{\psi}(x)(i\gamma\cdot\partial
+\gamma\cdot v+\gamma\cdot a\gamma_{5}
-mu(x))\psi(x)
+{1\over 2}m^{2}_{1}(\rho^{\mu}_{i}\rho_{\mu i}+
\omega^{\mu}\omega_{\mu}+a^{\mu}_{i}a_{\mu i}+f^{\mu}f_{\mu})
\nonumber \\
+{1\over 2}m^{2}_{2}(K^{*a}_{\mu}\bar{K}^
{*a\mu}+K_{1}^{\mu}K_{1\mu})
+{1\over 2}m^{2}_{3}(\phi_{\mu}\phi^{\mu}+f_{s}^{\mu}f_{s\mu})
\nonumber \\
+\bar{\psi}(x)_{L}\gamma\cdot W\psi(x)_{L}
+{\cal L}_{W}+{\cal L}_{lepton}-\bar{\psi}M\psi,
\end{eqnarray}
where \(a_{\mu}=\tau_{i}a^{i}_{\mu}+\lambda_{a}K^{a}_{1\mu}
+({2\over 3}+{1\over \sqrt{3}}\lambda_{8})
f_{\mu}+({1\over 3}-{1\over \sqrt{3}}\lambda_{8})
f_{s\mu}\)(\(i=1,2,3\) and \(a=4,5,6,7\)),
\(v_{\mu}=\tau_{i}
\rho^{i}_{\mu}+\lambda_{a}K^{*}_{\mu}+
({2\over 3}+{1\over \sqrt{3}}\lambda_{8})
\omega_{\mu}+({1\over 3}-{1\over \sqrt{3}}\lambda_{8})
\phi_{\mu}\),
$W^{i}_{\mu}$ is the W boson, and \(u=exp\{\gamma_{5}i
(\tau_{i}\pi_{i}+
\lambda_{a}K^{a}+\eta+
\eta')\}\), $m$ is a parameter, and M is the mass matrix of u, d, s quarks,
The masses $m^2_1$, $m^2_2$, and $m^2_3$ have been determined theoretically.
 
Using the notations of Ref.[1], we have
\begin{eqnarray}
\lefteqn{<\pi^i\pi^j|A_{\mu}|K>
=\frac{i}{m_{K}}\{(p_{1}+p_{2})_{\mu}F^{ij}
+(p_{1}-p_{2})_{\mu}G^{ij}
+q_{\mu}R^{ij}\},}\nonumber \\
&&<\pi^i\pi^j|V_{\mu}|K>=\frac{H^{ij}}
{m^{3}_{K}}\varepsilon^{\mu\nu\lambda\rho}p_{\nu}(p_{1}+p_{2})
_{\lambda}(p_{1}-p_{2})_{\rho},
\end{eqnarray}
where $p_{1}, p_{2}, p$ are momenta of two pions and kaon respectively,
\(q=p-p_1-p_2\), and \(i,j=+,-,0\).
We define
\[q^2_1=(p-p_1)^2,\;\;\;q^2_2=(p-p_2)^2,\;\;\;q^2_3=(p_1+p_2)^2.\]
The form factors, $F^{ij}, G^{ij}, R^{ij}$ and $H^{ij}$ are functions of
$q^2, q^2_1, q^2_2$, and $q^2_3$. These four variables satisfy
\[q^2_1+q^2_2+q^2_3=m^2_K+2m^2_\pi+q^2.\]
The paper is organized as: 1)introduction; 2)isospin relation;
3)form factors of vector current; 4)$K^*\rightarrow K\pi\pi$ decay;
5)form factors of
axial-vector current; 6)decay rates; 7)conclusions.
 
\section{Isospin relation}
For the decay modes $K^-\rightarrow\pi^+\pi^- l\nu, \pi^0\pi^0 l\nu$ and
$\bar{K^0}\rightarrow\pi^+\pi^0 l\nu$ there are isospin relations between
the form factors denoted as $A^{ij}$.
We take -$\pi^+$, $\pi^0$, and $\pi^-$ as isospin triplet and -$\bar{K^0}$
and $K^-$ as isospin doublet. The isospin relation is obtained as
\begin{equation}
A^{+-}=A^{00}-\frac{1}{\sqrt{2}}A^{+0},
\end{equation}
where \(A^{ij}=F^{ij}, G^{ij}, R^{ij}, H^{ij}\) respectively.
 
\section{Form factors of vector current}
The Vector Meson Dominance(VMD) is a natural result of this theory[3].
The coupling between the W bosons and the bosonized vector current($\Delta s=1\))
has been
derived as[5]
\begin{eqnarray}
{\cal L}^{V}&=&\frac{g_{W}}{4}sin\theta_{C}g\{
-{1\over2}(\partial_{\mu}W^{+}_{\nu}-\partial_{\nu}W^{+}_{\mu})
(\partial_{\mu}K^{*-}_{\nu}-\partial_{\nu}
K^{*-}_{\mu})
-{1\over2}(\partial_{\mu}W^{-}_{\nu}-\partial_{\nu}W^{-}_{\mu})
(\partial_{\mu}K^{*+}_{\nu}\\
&&-\partial_{\nu}
K^{*+}_{\mu})
+W^{+}_{\mu}j^{-}_{\mu}+W^{-}_{\mu}j^{+}_{\mu}\},
\end{eqnarray}
where $j^{\pm}_{\mu}$ is obtained by substituting
\[K^{\pm}_{\mu}\rightarrow{g_{W}\over4}sin\theta_{C}gW^{\pm}_{\mu}\]
into the vertex in which $K_{\mu}$ field is involved.
 
The matrix elements of the vector current of $K_{l4}$ are resulted in anomalous
vertices of mesons.
The two subprocesses are shown in Fig.1(a,b). There is contact term.
Three kinds of vertices are involved: the contact term $
{\cal L}_{K^{*}K\pi\pi}$, ${\cal L}_{K^{*}K^{*}\pi}$ and ${\cal L}_{K^{*}K\pi}$
, and ${\cal L}_{K^{*}K\rho}$ and ${\cal L}_{\rho\pi\pi}$.
In the chiral limit, $m_q\rightarrow 0$, all these vertices
have been derived from the Lagrangian (1)
[3] and are listed below
\begin{eqnarray}
\lefteqn{{\cal L}_{K^{*}K^{*}\pi}=-\frac{N_{C}}{\pi^{2}g^{2}f_{\pi}}\
\varepsilon^{\mu\nu\alpha\beta}d_{aci}K^{a}_{\mu}\partial_{\nu}K^{c}_{\alpha}
\partial_{\beta}\pi^{i}},\nonumber \\
&&{\cal L}_{K^{*}K\pi}={2\over g}f(q^{2})f_{abi}K^{a}_{\mu}(\partial_{\mu}
\pi^{i}K^{b}-\pi^{i}\partial_{\mu}K^{b}),\nonumber \\
&&f(q^{2})=1+\frac{q^{2}}{2\pi^{2}f^{2}_{\pi}}[(1-\frac{2c}{g})^{2}-4\pi^{2}
c^{2})],\nonumber \\
&&c=\frac{f^{2}_{\pi}}{2gm^{2}_{\rho}},\nonumber \\
&&{\cal L}_{K^{*}\rho K}=-\frac{N_{C}}{\pi^{2}g^{2}f^{2}_{\pi}}\varepsilon
^{\mu\nu\alpha\beta}d_{abi}K^{a}_{\mu}\partial_{\nu}\rho^{i}_{\alpha}
\partial_{\beta}K^{b},\nonumber \\
&&{\cal L}_{\rho\pi\pi}={2\over g}f(q^{2})\epsilon_{ijk}\rho^{i}_{\mu}
\pi^{j}\partial_{\mu}\pi^{k},\\
&&{\cal L}_{K^{*}K\pi\pi}=\frac{2}{g\pi^{2}f^{3}_{\pi}}(1-\frac{6c}{g}+
\frac{6c^{2}}{g^{2}})d_{abe}f_{cde}\varepsilon^{\mu\nu\alpha\beta}K^{a}_{\mu}
\partial_{\nu}P^{b}\partial_{\alpha}P^{c}\partial_{\beta}P^{d},
\end{eqnarray}
 
From theses vertices the form factors $h^{ij}$ are found
\begin{eqnarray}
\lefteqn{H^{+-}=\frac{m^{3}_{K}m^{2}_{K^{*}}}{q^{2}-m^{2}_{K^{*}}}
\{\frac{1}{\pi^{2}f^{3}_{\pi}}(1-\frac{6c}{g}
+\frac{6c^{2}}{g^{2}})-\frac{N_{C}}{g^{2}\pi^{2}f_{\pi}}
\frac{f(q^{2}_{2})}{q^{2}_{2}-m^{2}_{K^{*}}}}\nonumber \\
&&-\frac{N_{C}}{2g^{2}\pi^{2}f_{\pi}}\frac{f(q^{2}_{3})}{q^{2}_{3}-m^{2}_{\rho}
+i\sqrt{q^{2}_{3}}\Gamma(q^{2}_{3})}\},\\
&&H^{00}=-\frac{m^{3}_{K}m^{2}_{K^{*}}}{q^{2}-m^{2}_{K^{*}}}\frac{
N_{c}}{2g^{2}\pi^{2}f_{\pi}}\{\frac{f(q^{2}_{2})}{q^{2}
_{2}-m^{2}_{K^{*}}}-\frac{f^{2}(q^2_{1})}{q^{2}_{1}-m^{2}_{K^{*}}}\},\\
&&H^{+0}={1\over\sqrt{2}}\frac{m^{3}_{K}m^{2}_{K^{*}}}{q^{2}-m^{2}_{K^{*}}}\{
-\frac{2}{\pi^{2}f^{3}_{\pi}}(1-\frac{6c}{g}+\frac{6c^{2}}{g^{2}})+
\frac{N_{C}}{\pi^{2}g^{2}f_{\pi}}[\frac{f(q_{1})}{q^{2}_{1}-m^{2}_{K^{*}}}
+\frac{f(q_{2})}{q^{2}_{2}-m^{2}_{K^{*}}}\nonumber \\
&&+\frac{f(q_{3})}{q^{2}_{3}-m^{2}
_{\rho}+i\sqrt{q^{2}_{3}}\Gamma_{\rho}(q^{2}_{3})}]\},
\end{eqnarray}
where
$\Gamma_{\rho}$ is the decay width of $\rho$ meson
\begin{equation}
\Gamma_{\rho}(q^{2}_{3})=\frac{\sqrt{q^{2}_{3}}f^{2}(q^{2}_{3})}{12g^{2}\pi}
(1-\frac{4m^{2}_{\pi}}{q^{2}_{3}})^{{3\over2}}.
\end{equation}
The equations(8-10) show that the isospin relation(3)
is satisfied.
 
\section{$K^{*}\rightarrow K\pi\pi$ decay}
The vertices(6,7) are responsible for the decay of $K^{*}
\rightarrow K\pi\pi$. As a test the decay widths of $K^{*}\rightarrow K\pi\pi$
are calculated
\begin{equation}
\Gamma(K^{*-}\rightarrow K^{-}\pi^{+}\pi^{-})=\frac{1}{
96(2\pi)^{3}m_{K^{*}}}\int dk^{2}_{1}dk^{2}_{2}\{p^{2}_{1}p^{2}_{2}-
(\vec{p}_{1}\cdot\vec{p}_{2})^{2}\}|A|^{2}
=0.29\times10^{-5}GeV
\end{equation}
which is less than the experimental upper limit[9],
where A is the amplitude
\begin{eqnarray}
A&=&\frac{4}{g\pi^{2}f^{3}_{\pi}}(1-\frac{6c}{g}+\frac{6c^{2}}{g^{2}})
-\frac{4N_{c}}{g^{3}\pi^{2}f_{\pi}}\frac{f(k_{2}^{2})}{k^{2}_{2}
-m^{2}_{K^{*}
}+i\sqrt{k^{2}_{2}}\Gamma_{K^{*}}(k^{2}_{2})}\nonumber \\
&&-\frac{2N_{c}}{g^3\pi^2f_\pi}\frac{f(k^{2}_{3})}
{k^{2}_{3}-m^{2}_{\rho}+i\sqrt{k^{2}_{3}}\Gamma_{\rho}(k^{2}_{3})}
\end{eqnarray}
where \(k^{2}_{1}=(p+p_{1})^{2}\), \(k^{2}_{2}=(p+p_{2})^{2}\),
\(k^{2}_{3}
=(p_{1}+p_{2})^{2}\), and $p_{1}, p_{2}, p$ are momenta of $\pi^{+},
\pi^{-}$ and $K^{-}$ respectively,
$\Gamma_{K^{*}}$ is the decay width of $K^{*}$
\begin{equation}
\Gamma_{K^{*}}(k^{2}_{2})=\frac{f^2(k^{2}_{2})}{2\pi g^{2}k^{2}_{2}}
\{{1\over4k^{2}_{2}}(k^{2}_{2}+m^{2}_{K}-m^{2}_{\pi})^{2}-m^{2}_{K}\}
^{{3\over2}}.
\end{equation}
\begin{eqnarray}
\lefteqn{\Gamma(K^{*-}\rightarrow K^{-}\pi^{0}\pi^{0})=\frac{1}{
192(2\pi)^{3}m_{K^{*}}}\int dk^{2}_{1}dk^{2}_{2}\{p^{2}_{1}p^{2}_{2}-
(\vec{p}_{1}\cdot\vec{p}_{2})^{2}\}}\nonumber \\
&&\frac{36}{\pi^{4}g^{6}f^{2}_{\pi}}
\{\frac{f(k_{1})}{k^{2}_{1}-m^{2}_{K^{*}}+i\sqrt{k^{2}_{1}}\Gamma_{K^{*}}
(k^{2}_{1})}-
\frac{f(k_{2})}{k^{2}_{2}-m^{2}_{K^{*}}+i\sqrt{k^{2}_{2}}\Gamma_{K^{*}}
(k^{2}_{2})}\}^{2}\nonumber \\
&&=0.61\times10^{-6}GeV.
\end{eqnarray}
\begin{equation}
\Gamma(K^{*-}\rightarrow\bar{K}^{0}\pi^{-}\pi^{0})=\frac{1}{
96(2\pi)^{3}m_{K^{*}}}\int dk^{2}_{1}dk^{2}_{2}\{p^{2}_{1}p^{2}_{2}-
(\vec{p}_{1}\cdot\vec{p}_{2})^{2}\}|B|^{2}=0.38\times10^{-4}GeV,
\end{equation}
where
\begin{eqnarray}
\lefteqn{B=-\frac{8}{\sqrt{2}gf^{3}_{\pi}}(1-\frac{6c}{g}+\frac{6c^{2}}{g^{2}})
+\frac{12}{\sqrt{2}\pi^{2}g^{3}f_{\pi}}\{
\frac{f(k_{1})}{k^{2}_{1}-m^{2}_{K
^{*}}+i\sqrt{k^{2}_{1}}\Gamma_{K^{*}}(k^{2}_{1})}}\nonumber\\
&&+\frac{f(k_{2})}{k^{2}_{2}-m^{2}_{K
^{*}}+i\sqrt{k^{2}_{2}}\Gamma_{K^{*}}(k^{2}_{2})}+
\frac{f(k_{3})}{k^{2}_{3}-m^{2}_{\rho
}+i\sqrt{k^{2}_{3}}\Gamma_{\rho}(k^{2}_{3})}\}.
\end{eqnarray}
Eq.(16) is compatible with the data
9Y.
 
\section{Form factors of axial-vector current}
In the chiral limit, the axial-vector part of the interaction between W-boson
and mesons is expressed as[5]
\begin{eqnarray}
\lefteqn{{\cal L}^{As}=
{g_{W}\over 4}{1\over f_{a}}
sin\theta_{C}\{-{1\over 2}(\partial_{\mu}W^{\pm}
_{\nu}-\partial_{\nu}W^{\pm}_{\mu})
(\partial^{\mu}K^{\mp
\nu}_{a}-\partial^{\nu}K^{\mp\mu}_{a})+W^{\pm\mu}j^{\mp}_{\mu}\}}
\nonumber \\
&&+{g_{W}\over 4}sin\theta_{C}
\Delta m^{2}f_{a}W^{\pm}_{\mu}K^{\mp\mu}
+{g_w\over4}\sin\theta_{C}
f_{K}W^{\pm}_{\mu}\partial^{\mu}K^{\mp},
\end{eqnarray}
where $j^{\pm}_{\mu}$ are obtained by substituting $K^{\pm}_{a\mu}
\rightarrow{g_{W}\over4f_{a}}
sin\theta_{C}W^{\pm}_{\mu}$ into the vertex in
which $K_{a}$ fields are involved,
\begin{eqnarray}
\lefteqn{f_{a}=g^{-1}(1-{1\over2\pi^{2}g^{2}})^{-{1\over2}},}\\
&&\Delta m^{2}=6m^{2}g^{2}=
f^{2}_{\pi}(1-{f^{2}_{\pi}\over g^{2}m^{2}_{\rho}})^{-1},\\
&&c=\frac{f^2_{\pi}}{2gm^2_{\rho}}.
\end{eqnarray}
The mass of $K_1$ meson is determined by
\begin{equation}
(1-{1\over 2\pi^{2}g^{2}})m^{2}_{K_1}=6m^{2}+m^{2}_{K^*}.
\end{equation}
The numerical value is \(m_{k_1}=1.322GeV\) which is compatible with the data
[9].
 
Two subprocesses contribute to the matrix element of the axial-vector current.
They are shown in Fig.2(a,b). The vertices of mesons involved in these processes
are ${\cal L}_{K_1 K^*\pi}$, ${\cal L}_{K^* K\pi}$ and
${\cal L}_{K_1 \rho K}$, ${\cal L}_{\rho\pi\pi}$. There is a contact term
${\cal L}_{K_1 K\pi\pi}$ too.
However, the calculation shows that the contribution
of the contact term is very small and negligible.
In the chiral limit, these vertices have been
derived from the Lagrangian(1)
\begin{eqnarray}
\lefteqn{{\cal L}_{K_1 K^*\pi}=f_{abi}\{A(p^2)K^a_{1\mu}K^{*b}_{\mu}\pi^i
-BK^a_{1\mu}K^{*b}_\nu \partial_{\mu\nu}\pi^i
+DK^a_{1\mu}\partial^\mu (K^{*b}_{\nu}\partial^{\nu}\pi^i)\}},\\
&&{\cal L}_{K_1 \rho K}=-f_{abi}\{A(p^2)K^a_{1\mu}\rho^{i}_{\mu}K^b
-BK^a_{1\mu}\rho^{i}_\nu \partial_{\mu\nu}K^b
+DK^a_{1\mu}\partial^{\mu}(\rho^{i}_{\nu}\partial^\nu K^b)\},
\end{eqnarray}
where
\begin{eqnarray}
\lefteqn{A(p^2)={2\over f_{\pi}}gf_{a}\{{F^2\over g^2}
+p^{2}[{2c\over g}+{3\over4
\pi^{2}g^{2}}(1-{2c\over g})]}\nonumber \\
&&+q^{2}[{1\over 2\pi^{2}g^{2}}-
{2c\over g}-{3\over4\pi^{2}g^{2}}(1-{2c\over g})]\},\\
&&F^2=f^2_\pi (1-{2c\over g})^{-1},\\
&&B=-{2\over f_{\pi}}gf_{a}{1\over2\pi^{2}g^{2}}(1-{2c\over g}),\\
&&D=-{2\over f_{\pi}}f_{a}\{2c+{3\over2\pi^{2}g}(1-{2c\over g})\},
\end{eqnarray}
where q and p are the momentum of $K_1$ and the vector meson respectively.
\begin{eqnarray}
\lefteqn{{\cal L}_{K^* K\pi}={2\over g}f_{abi}f(p^2)K^a_\mu (K^b\partial_
{\mu}\pi^i-\pi^i\partial_\mu K^b),}\\
&&{\cal L}_{\rho\pi\pi}={2\over g}\epsilon_{ijk}f(p^2)\rho^i_\mu\pi^j\partial
_\mu \pi^k,\\
&&f(p^2)=1+\frac{p^2}{2\pi f^2_\pi}[(1-{2c\over g})^2-4\pi^2 c^2],
\end{eqnarray}
where p is the momentum of the vector meson.
 
By using Eqs.(18,23,24), we obtain
\begin{eqnarray}
<\pi^+\pi^-|A_{\mu}|K^->&=&{1\over\sqrt{2}}
(\frac{q_\mu q_\nu}{q^2}-g_{\mu\nu})\frac{g^2f_a m^2_{K^*}}{q^2-m^2_{K_1}}
<\pi^+\pi^-|\{A(p_{K^*})\bar{K^0}_\nu \pi^- -B\bar{K^0}_\lambda
\partial_{\lambda\nu}\pi^-\}\nonumber \\
&&-{1\over\sqrt{2}}\{A(p_{\rho})\rho^0_\nu K^
- -B\rho^0_{\lambda}\partial_{\lambda\nu}K^-\}|K^->.
\end{eqnarray}
In the chiral limit PCAC is satisfied.
The reason is that the Lagrangian(1) is chiral symmetric in the limit
$m_q\rightarrow 0$. On the other hand, the satisfaction of PCAC is resulted
in the cancellations between the four terms of Eq.(18).
The Eq.(18) shows that the axial-vector current has more complicated structure
than the vector current does(5).
Because of the PCAC the form factor R(2) is not an independent quantity and
determined as
\begin{equation}
R=-{1\over q^2}\{q\cdot(p_1+p_2)F+q\cdot(p_1-p_2)G\}.
\end{equation}
Substituting the
vertices(29,30) into Eq.(32), the three form factors are obtained
\begin{eqnarray}
\lefteqn{F^{+-}=\frac{gf_a m^2_{K^*}m_K}{q^2-m^2_{K_1}}\{\frac{f(q^2_2)}
{q^2_2-m^2_{K^*}}[{3\over2}A(q^2_2)+{1\over2}Bp_1\cdot(p+p_2)]}\nonumber \\
&&+\frac{f(q^2_3)}{q^2_3-m^2_\rho+i\sqrt{q^2_3}\Gamma_{\rho}(q^2_3)}B
p\cdot(p_2-p_1)\},\\
&&G^{+-}=\frac{gf_a m^2_{K^*}m_K}{q^2-m^2_{K_1}}\{\frac{f(q^2_2)}
{q^2_2-m^2_{K^*}}[-{1\over2}A(q^2_2)+{1\over2}Bp_1\cdot(p+p_2)]\nonumber\\
&&-\frac{f(q^2_3)}{q^2_3-m^2_\rho+i\sqrt{q^2_3}\Gamma_{\rho}(q^2_3)}
A(q^2_3)\}.
\end{eqnarray}
 
In the same way the form factors of other two decay modes are obtained
\begin{eqnarray}
\lefteqn{F^{00}={1\over2}\frac{gf_a m^2_{K^*}m_K}{q^2-m^2_{K_1}}\{\frac{f(q^2_1)}
{q^2_1-m^2_{K^*}}[{3\over2}A(q^2_1)+{1\over2}B(p_2\cdot p+p_2\cdot p_1)]}
\nonumber\\
&&+\frac{f(q^2_2)}{q^2_2-m^2_{K^*}}[{3\over2}A(q^2_2)+{1\over2}B(p_1\cdot p+
p_1\cdot p_2)]\},\\
&&G^{00}={1\over2}\frac{gf_a m^2_{K^*}m_K}{q^2-m^2_{K_1}}\{\frac{f(q^2_1)}
{q^2_1-m^2_{K^*}}[{1\over2}A(q^2_1)-{1\over2}B(p_2\cdot p+p_2\cdot p_1)]
\nonumber \\
&&+\frac{f(q^2_2)}{q^2_2-m^2_{K^*}}[-{1\over2}A(q^2_2)+{1\over2}B(p_1\cdot p+
p_1\cdot p_2)]\},\\
&&F^{+0}={1\over\sqrt{2}}\frac{gf_a m^2_{K^*}m_K}{q^2-m^2_{K_1}}
\{\frac{f(q^2_1)}
{q^2_1-m^2_{K^*}}[{3\over2}A(q^2_1)+{1\over2}B(p_2\cdot p+p_2\cdot p_1)]
\nonumber\\
&&-\frac{f(q^2_2)}{q^2_2-m^2_{K^*}}[{3\over2}A(q^2_2)+{1\over2}B(p_1\cdot p+
p_1\cdot p_2)]\nonumber\\
&&+\frac{2f(q^2_3)}{q^2_3-m^2_\rho+i\sqrt{q^2_3}\Gamma_{\rho}(q^2_3)}B
p\cdot(p_1-p_2)\},\\
&&G^{+0}={1\over\sqrt{2}}\frac{gf_a M^2_{K^*}m_K}
{q^2-m^2_{K_1}}\{\frac{f(q^2_1)}
{q^2_1-m^2_{K^*}}[{1\over2}A(q^2_1)-{1\over2}B(p_2\cdot p+p_2\cdot p_1)]
\nonumber\\
&&-\frac{f(q^2_2)}{q^2_2-m^2_{K^*}}[-{1\over2}A(q^2_2)+{1\over2}B(p_1\cdot p+
p_1\cdot p_2)]
+\frac{2f(q^2_3)}{q^2_3-m^2_\rho+i\sqrt{q^2_3}\Gamma_{\rho}(q^2_3)}
A(q^2_3)\}.
\end{eqnarray}
The isospin relations(3) between these form factors are satisfied.
 
The partial wave analysis of these form factors can be done.
The decay channel $\rho\rightarrow \pi\pi$ contributes to the decay modes
of $\pi^+\pi^-$ and $\pi^+\pi^0$. The range of the variable $q^2_3$
is $4m^2_\pi<q^2_3<(m_K-m_l)^2$ in which the decay width $\Gamma_\rho(q^2_3)$
is not zero.
The form factors, $A^{+-}$ and $A^{+0}$ are complex functions of $q^2_3$.
The $\rho\rightarrow\pi\pi$ doesn't contribute to $\pi^0\pi^0$ mode. Therefore,
$F^{00}$ and $G^{00}$ are real.
$K_{l4}$ are decays at low energies. s- and p- waves are major partial waves.
The $q^2_1$ and $q^2_2$ variables are expressed as
\begin{eqnarray}
\lefteqn{q^2_1={1\over2}(m^2_K+2m^2_\pi+q^2-q^2_3)+(1-{4m^2_\pi\over q^2_3})
^{{1\over2}}Xcos\theta_\pi,}\\
&&q^2_2={1\over2}(m^2_K+2m^2_\pi+q^2-q^2_3)-(1-{4m^2_\pi\over q^2_3})
^{{1\over2}}Xcos\theta_\pi,
\end{eqnarray}
where \(X=\{{1\over4}(m^2_K-q^2-q^2_3)^2-q^2q^2_3\}^{{1\over2}}\) and $\theta
_\pi$ is the angle between $\vec{p}_1$ and $\vec{p}$ in the rest frame of the
two pions.
 
The s- and p- wave amplitudes are obtained from Eqs.(34-39)
\begin{enumerate}
\item $F^{+-}_s$ is real. Only Fig.2(a) contributes to it.
$F^{+-}_p$ is a complex function of $q^2_3$ resulted by $\rho\rightarrow
\pi\pi$. $F^{+-}_p$ has a phase shift.
\begin{equation}
F^{+-}=F^{+-}_s +|F^{+-}_p|e^{i\delta^{+-}_p}(1-{4m^{2}_\pi\over q^2_3})
^{{1\over2}}{X\over m^2_K}cos\theta_\pi.
\end{equation}
\item $G^{+-}_{s}$ is complex and has a phase shift. $G^{+-}_p$ is real.
\begin{equation}
G^{+-}=|G^{+-}_s|e^{i\delta^{+-}_s}+G^{+-}_p(1-{4m^2_\pi\over q^2_3})
^{{1\over2}}{X\over m^{2}_K}cos\theta_\pi.
\end{equation}
\item Both $G^{00}_s$ and $G^{00}_p$ are real.
\begin{eqnarray}
F^{00}&=&F^{00}_s,\nonumber \\
G^{00}&=&|G^{00}_p|(1-{4m^2_\pi\over q^2_3})^{{1\over2}}{X\over m^2_K}cos\theta_\pi.
\end{eqnarray}
\item The isospin of the two pions of the $\pi^+\pi^0$ mode is one.
Because of Bose statistics $F^{+0}$ only has p-wave which is complex and
has phase shift.
$G^{+0}$ has s wave only. $G^{+0}_s$ is complex and it has phase shift.
\begin{eqnarray}
F^{+0}&=&|F^{+0}_p|e^{i\delta^{+0}_p}(1-{4m^2_\pi\over q^2_3})^{{1\over2}}
{X\over m^2_K}cos\theta_\pi,
\nonumber \\
G^{+0}&=&|G^{+0}_s|e^{\delta^{+0}_s}.
\end{eqnarray}
\end{enumerate}
All the phase shifts are caused by the decay $\rho\rightarrow\pi\pi$ and functions of $q^2$ and $q^2_3$.
 
\section{Decay rates}
The decay rates of the three modes of $K_{e4}$ and $K_{\mu4}$ are calculated.
As mentioned above, all the form factors are derived in the chiral limit.
Therefore, only the leading terms of the masses of kaon and pions are kept
in the calculation of the decay rates.
 
Ignoring $m_e$, only the form factors F, G, and H contribute to the decay
rates of $K_{e4}$.
By using the formula of Ref.[1] we obtain
\[\Gamma(K^-\rightarrow\pi^+\pi^- e\nu)=2.06\times10^{-21}GeV,\;\;\;
B=3.87\times10^{-5}.\]
\[\Gamma(K^-\rightarrow\pi^0\pi^0 e\nu)=0.221\times10^{-21}GeV,\;\;\;
B=0.42\times10^{-5}.\]
\[\Gamma(K^-\rightarrow\pi^+\pi^0 e\nu)=3.24\times10^{-21}GeV,\;\;\;
B=2.55\times10^{-4}.\]
The experimental data are
\[B(\pi^{+}\pi^-)=(3.91\pm0.17)\times10^{-5}[10],\]
\[B(\pi^{0}\pi^0)=(2.54\pm0.89)\times10^{-5}(10\; events)[11],\]
\[B(\pi^{-}\pi^0)=(5.16\pm0.20\pm0.22)\times10^{-5}[12],\]
\[B(\pi^{-}\pi^0)=(6.2\pm2.0)\times10^{-5}[13],\]
\[B(\pi^{-}\pi^0)<200\times10^{-5}[14].\]
Theoretical result of $\pi^+\pi^-$ mode agrees well with the data.

The form factors of the vector current are determined by anomalous vertices.
The numerical calculation shows that
the contribution of the form factor H is
only $0.5\%$ of the total decay rate of $K^{-}\rightarrow\pi^+\pi^- e\nu$.
Therefore, the axial-vector current dominates the $K_{l4}$ decays.
 
As shown in Fig.2(a,b) there are two channels in $K_{l4}$ decays. Numerical
calculation of $K^- \rightarrow\pi^+\pi^- e\nu$
shows that the contribution of $\rho\rightarrow\pi\pi$(Fig.2(b)) is twice of
the process, $K^*\rightarrow K\pi$, (Fig.2(a)). Only the process(Fig.2(a)) contribute to
$K^-\rightarrow\pi^0\pi^0 e\nu$. Because of Bose statistics
there is an additional factor of ${1\over2}$ in the formula of the decay rate of
this mode. Therefore, this theory predicts
 smaller decay rate for this decay
mode.
On the other hand, the numerical calculation shows that the process(Fig.2(b))
is the major contributor of the decay $\bar{K^0}\rightarrow\pi^+\pi^0 e\nu$.
The theory predicts a larger branching ratio for $\bar{K}^0\rightarrow\pi^{+}\pi^0 e\nu$.\\
All the form factors contribute to $K_{\mu 4}$ decays. Eq.(33) shows that in
the chiral limit PCAC predicts that the form factor R is determined by
other two form factors, F and G. The branching ratio of $K_{\mu 4}$
provides a test on this prediction.
The numerical results are
\[\Gamma(K^-\rightarrow\pi^+\pi^- \mu\nu)=0.634\times10^{-21}GeV,\;\;\;
B=1.19\times10^{-5}.\]
\[\Gamma(K^-\rightarrow\pi^0\pi^0 \mu\nu)=0.673\times10^{-22}GeV,\;\;\;
B=0.126\times10^{-5}.\]
\[\Gamma(\bar{K}^0\rightarrow\pi^+\pi^0 \mu\nu)=1.01\times10^{-21}GeV,\;\;\;
B=0.793\times10^{-4}.\]
The experimental data[9] is
\[B(K^{-}\rightarrow\pi^+\pi^-\mu\nu)=(1.4\pm0.9)\times10^{-5}.\]
Theory agrees with the data well.
 
\section{Conclusion}
All the four form factors of $K_{l4}$
have been derived from an effective theory of large $N_{C}$
QCD in the chiral limit. It has been found that the contribution of
the vector current is negligible and the axial-vector current is dominant in
$K_{l4}$ decays. PCAC is revealed from the theory. In the chiral limit it has been predicted that the form factor R is determined by the form factors F and G.
The prediction has been tested by $K^-\rightarrow\pi^+\pi^-\mu\nu$. Theory
agrees with the data. The partial wave analysis has been done.
Non-zero phase shifts originate in the decay $\rho\rightarrow\pi\pi$. The
process $K_1\rightarrow\rho K$ and $\rho\rightarrow\pi\pi$(Fig.2(b)) plays
important role in $K_{l4}$ decays. Because of this channel the theory
predicts larger branching ratio for $K^-\rightarrow\pi^+\pi^- e\nu$ and
$\bar{K}^0\rightarrow\pi^+\pi^0 e\nu$. The former agrees well with the data.
$\rho$ resonance doesn't contribute to
$K^- \rightarrow\pi^0\pi^0 e\nu$. Therefore, the branching ratio of this decay
mode is predicted to be smaller.
 
This research was partially
supported by DOE Grant No. DE-91ER75661.

\pagebreak
\begin{flushleft}
{\bf Figure Captions}
\end{flushleft}
{\bf Fig.1} Feynman Diagrams of vector current
\\{\bf Fig.2}
 Feynman diagrams of axial-vector current
\\{\bf Fig.3}
Phase shifts
\\{\bf Fig.4}
Phase shifts
{\bf Fig.5} Phase shifts
\\{\bf Fig.5}
Phase shifts
\\{\bf Fig.6}
Phase shifts
\\{\bf Fig.7}
Form factors od $\pi^+\pi^-$ mode
\\{\bf Fig.8}
Form factors od $\pi^+\pi^-$ mode
\\{\bf Fig.9}
Form factors od $\pi^+\pi^0$ mode
\\{\bf Fig.10}
Form factors od $\pi^+\pi^0$ mode
\\{\bf Fig.11}
Form factors od $\pi^0\pi^0$ mode
\\{\bf Fig.12}
Form factors od $\pi^0\pi^0$ mode
\begin{figure}
\begin{center}
\includegraphics[width=7in, height=7in]{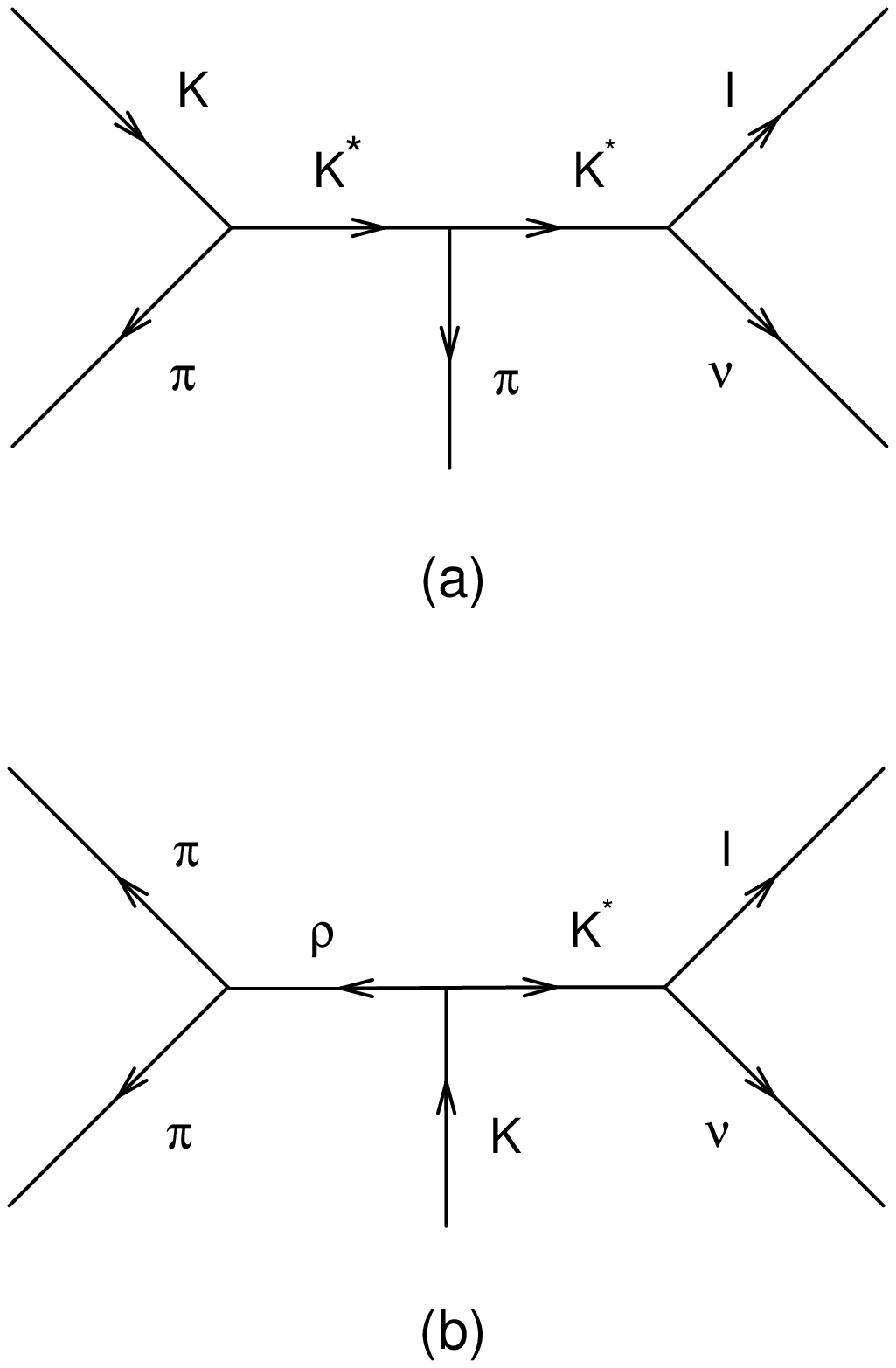}
Fig.1
\end{center}
\end{figure}
 
\begin{figure}
\begin{center}
\includegraphics[width=7in, height=7in]{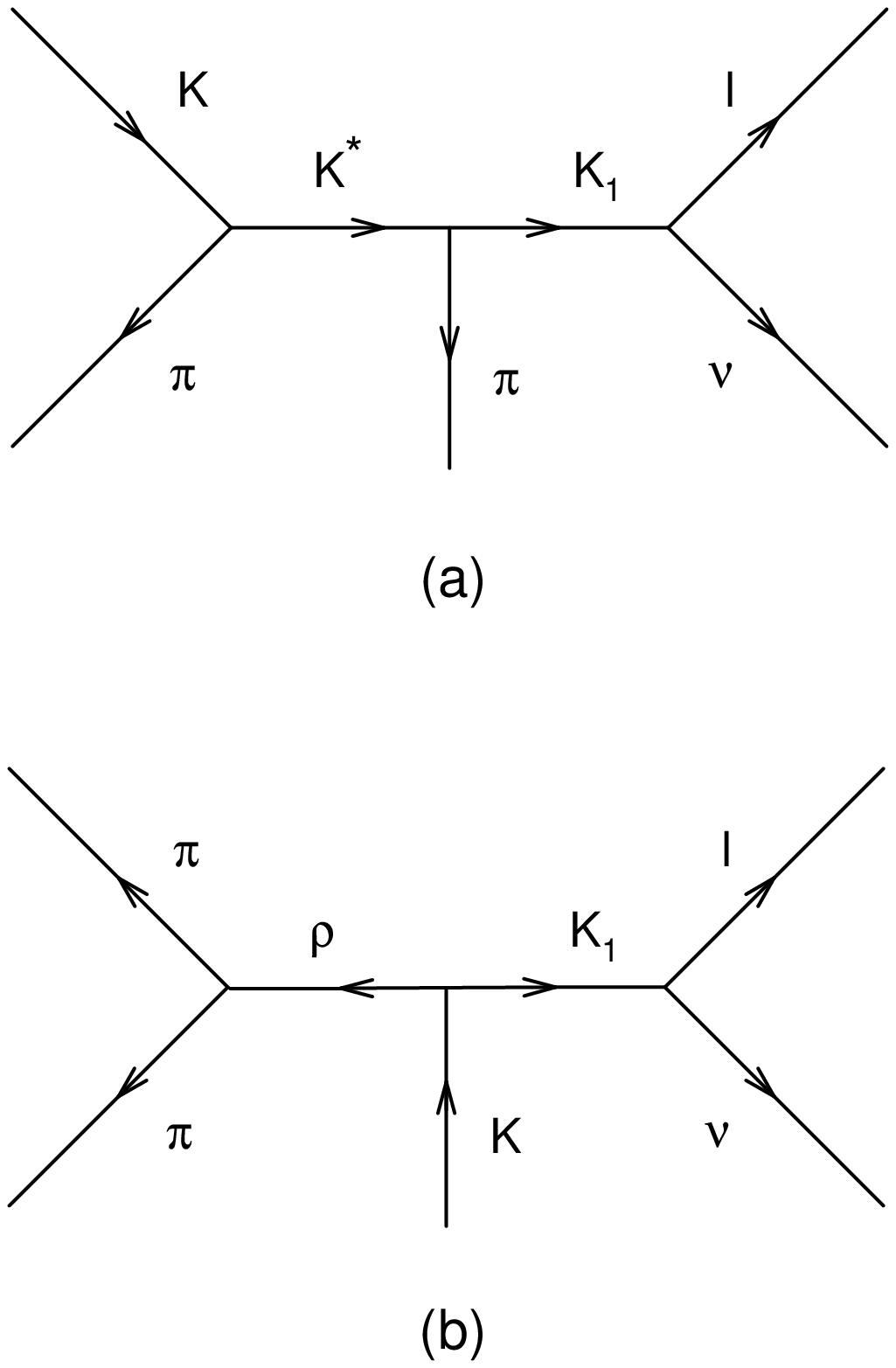}
Fig.2
\end{center}
\end{figure}
\begin{figure}
\begin{center}
\includegraphics[width=7in, height=7in]{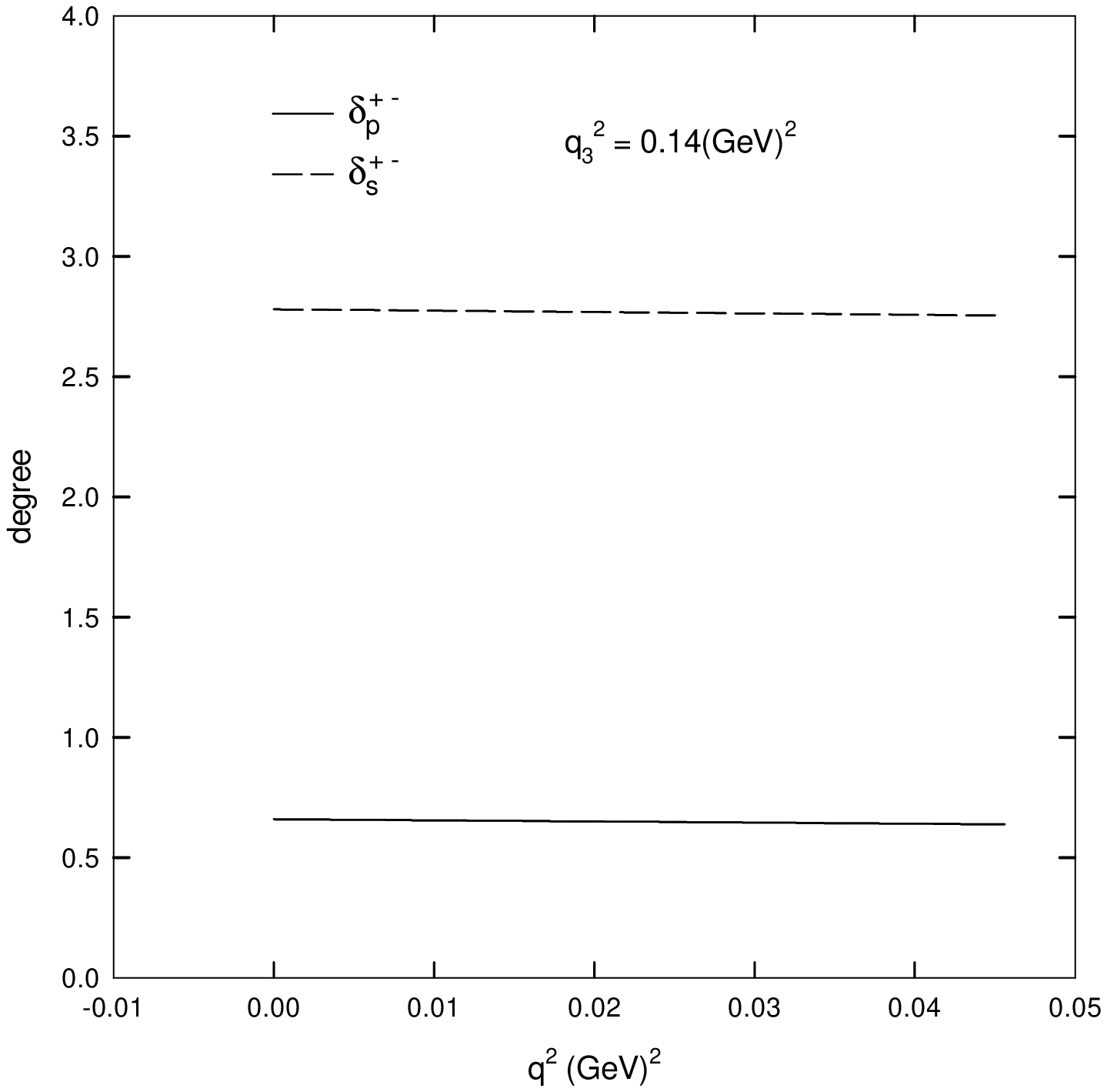}
Fig.3
\end{center}
\end{figure}
 
\begin{figure}
\begin{center}
\includegraphics[width=7in, height=7in]{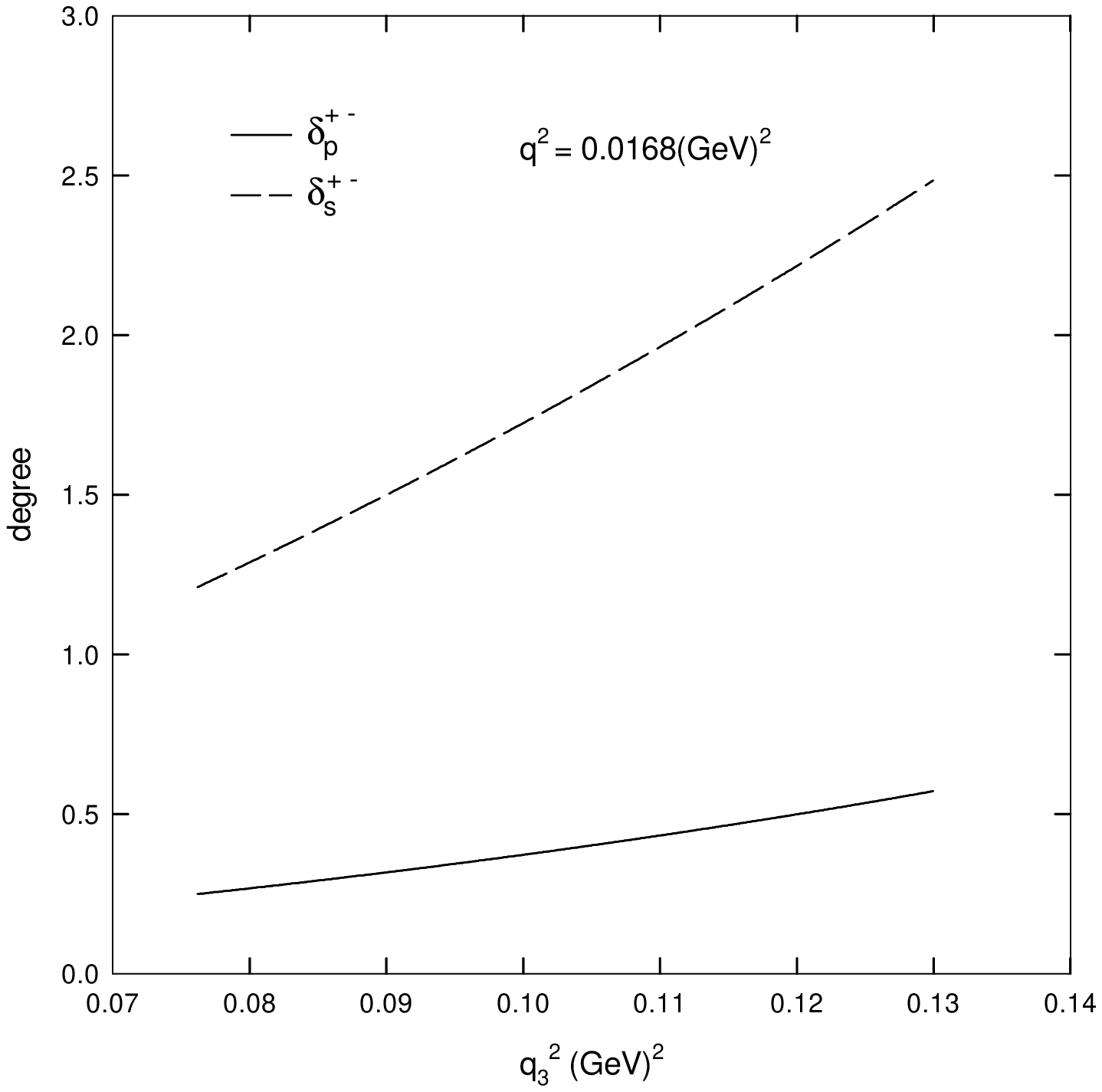}
Fig.4
\end{center}
\end{figure}
 
\begin{figure}
\begin{center}
\includegraphics[width=7in, height=7in]{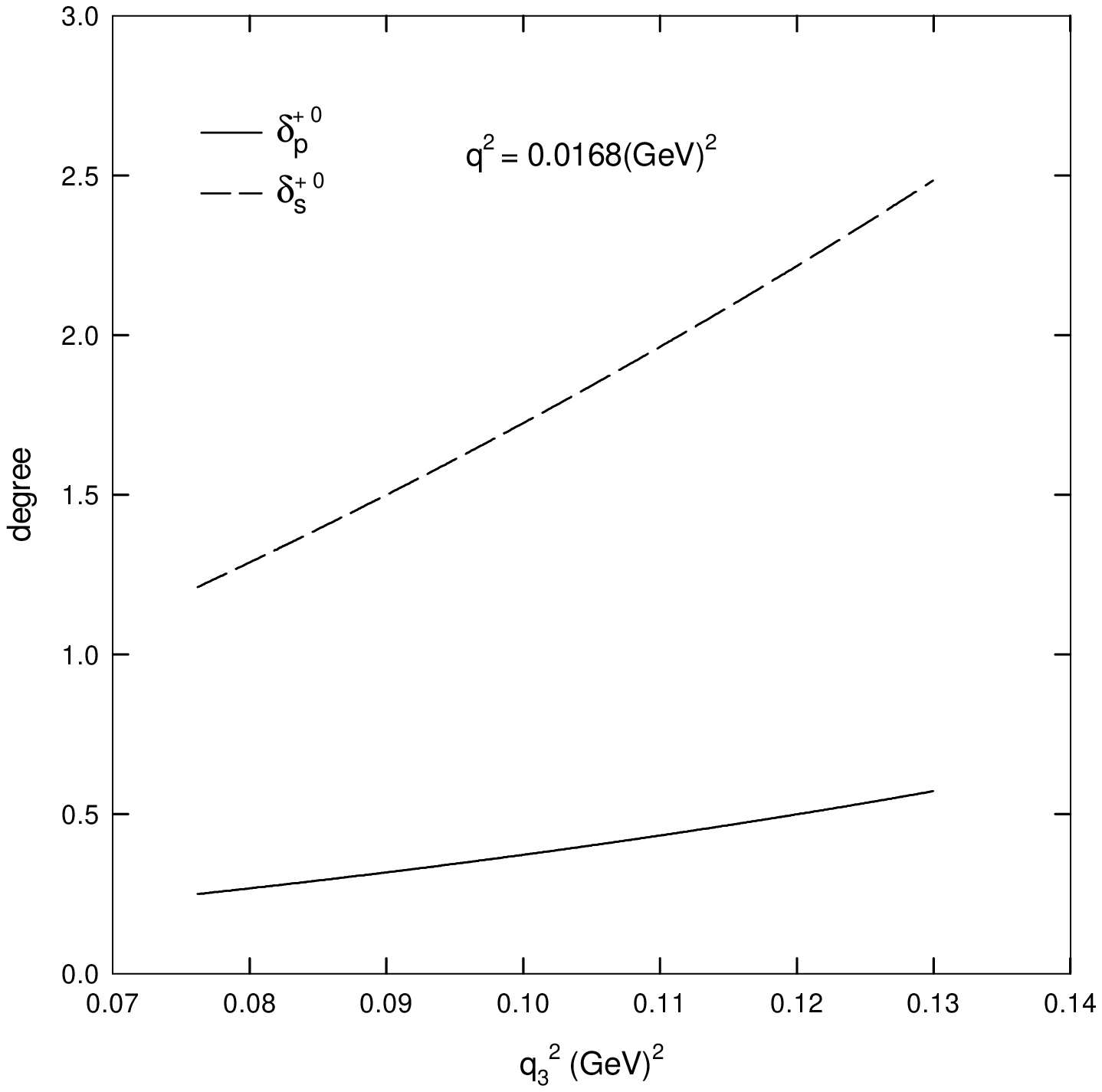}
Fig.5
\end{center}
\end{figure}
 
\begin{figure}
\begin{center}
\includegraphics[width=7in, height=7in]{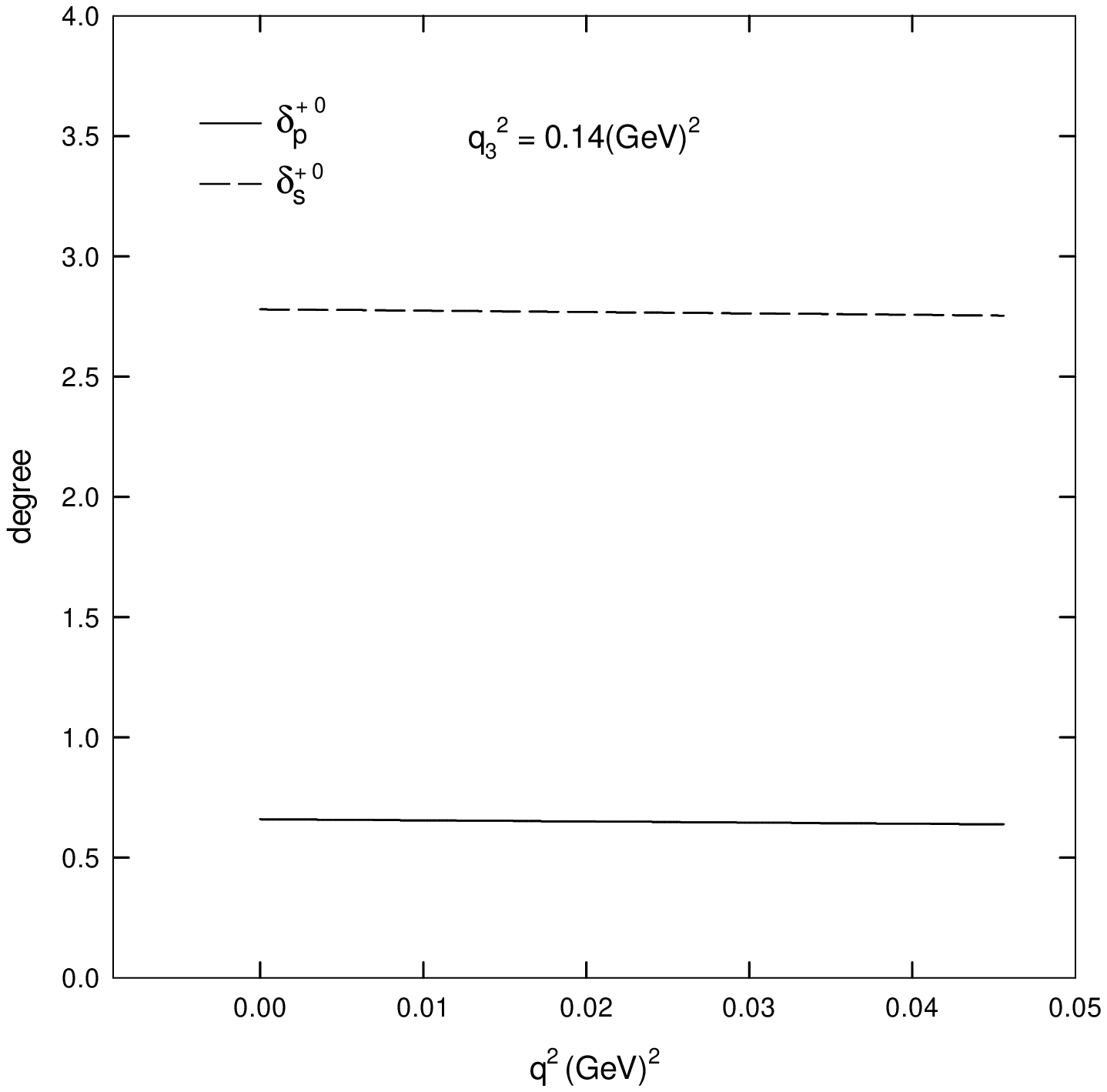}
Fig.6
\end{center}
\end{figure}
 
\begin{figure}
\begin{center}
\includegraphics[width=7in, height=7in]{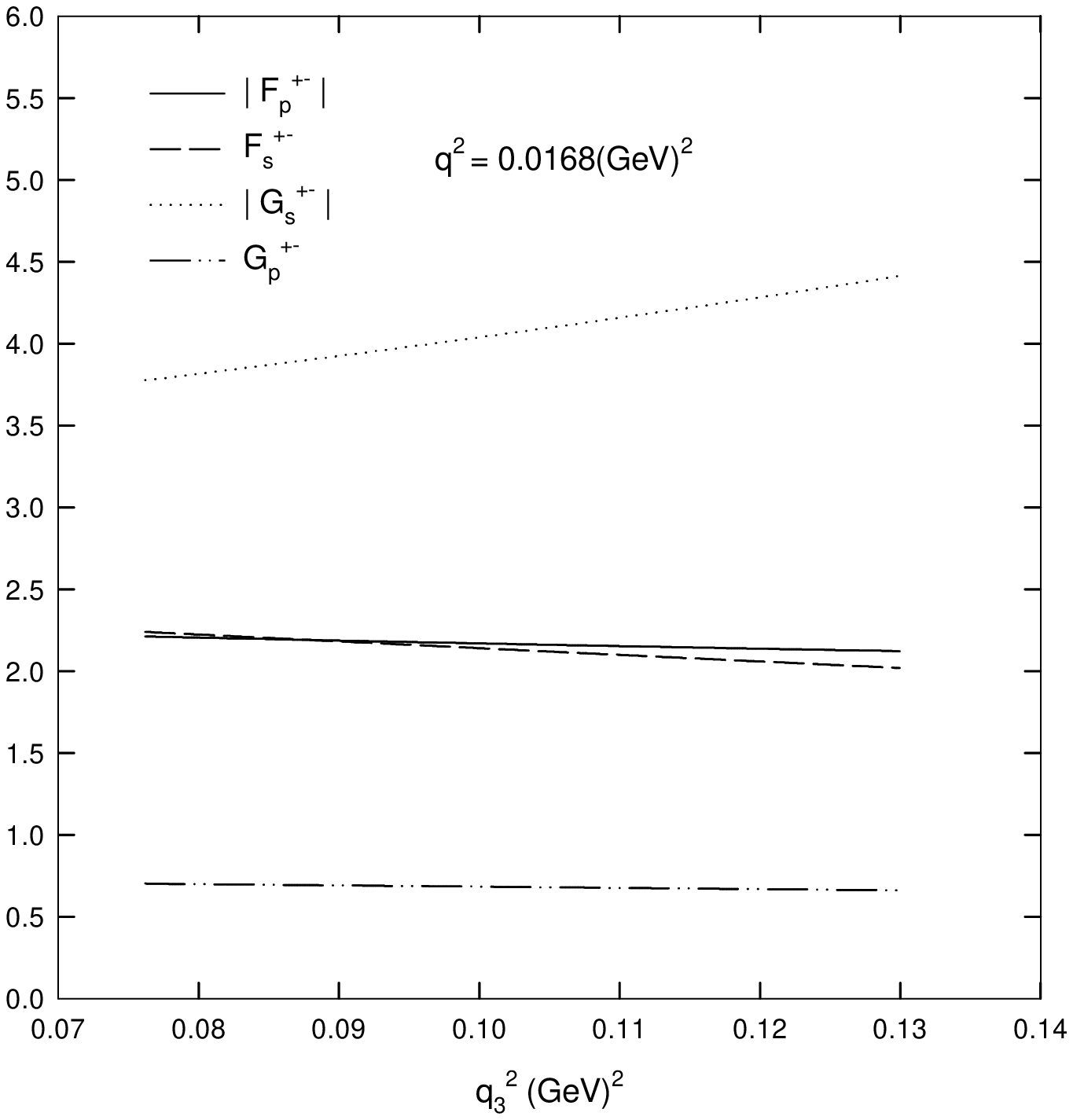}
Fig.7
\end{center}
\end{figure}
 
\begin{figure}
\begin{center}
\includegraphics[width=7in, height=7in]{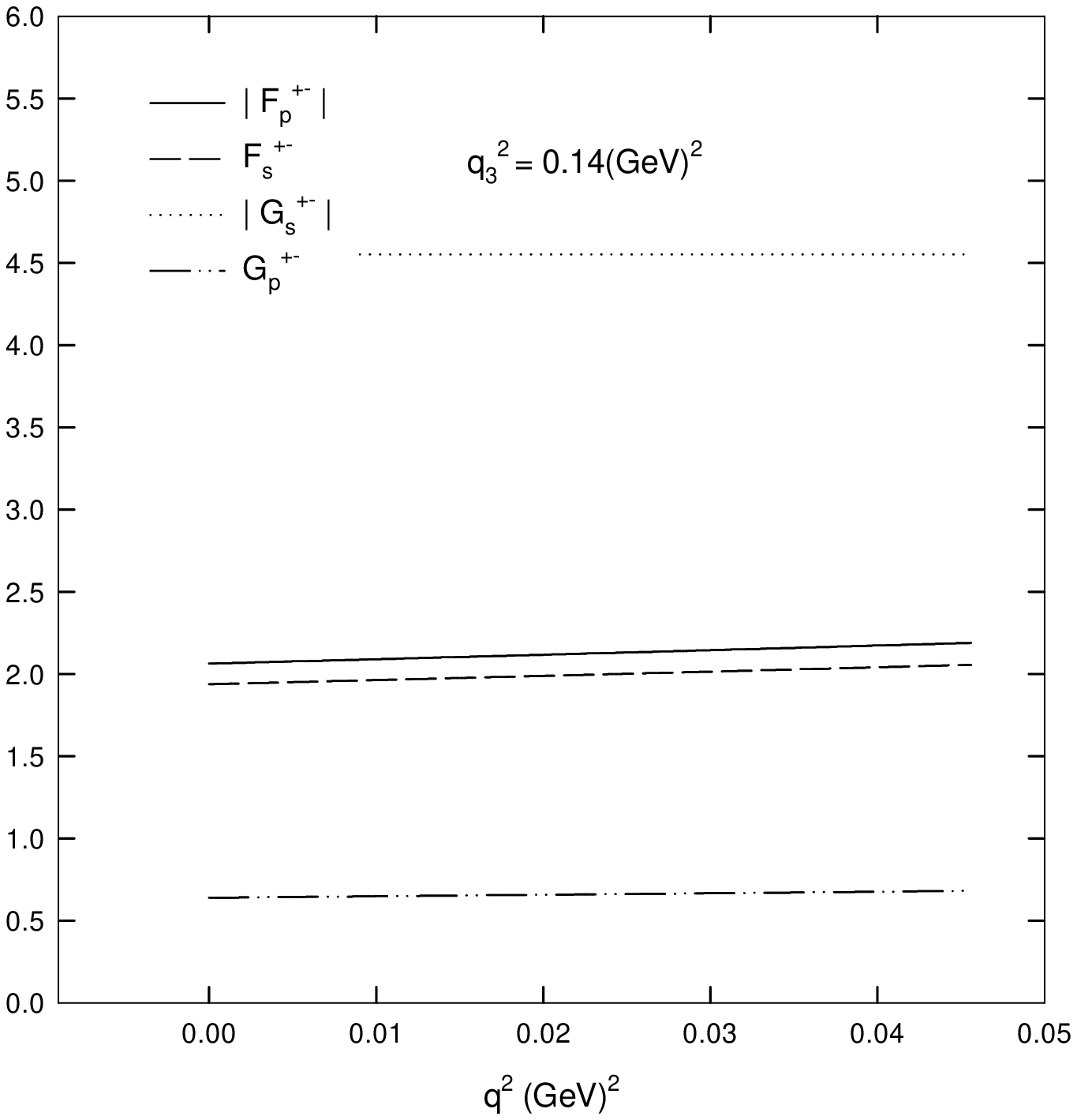}
Fig.8
\end{center}
\end{figure}
 
\begin{figure}
\begin{center}
\includegraphics[width=7in, height=7in]{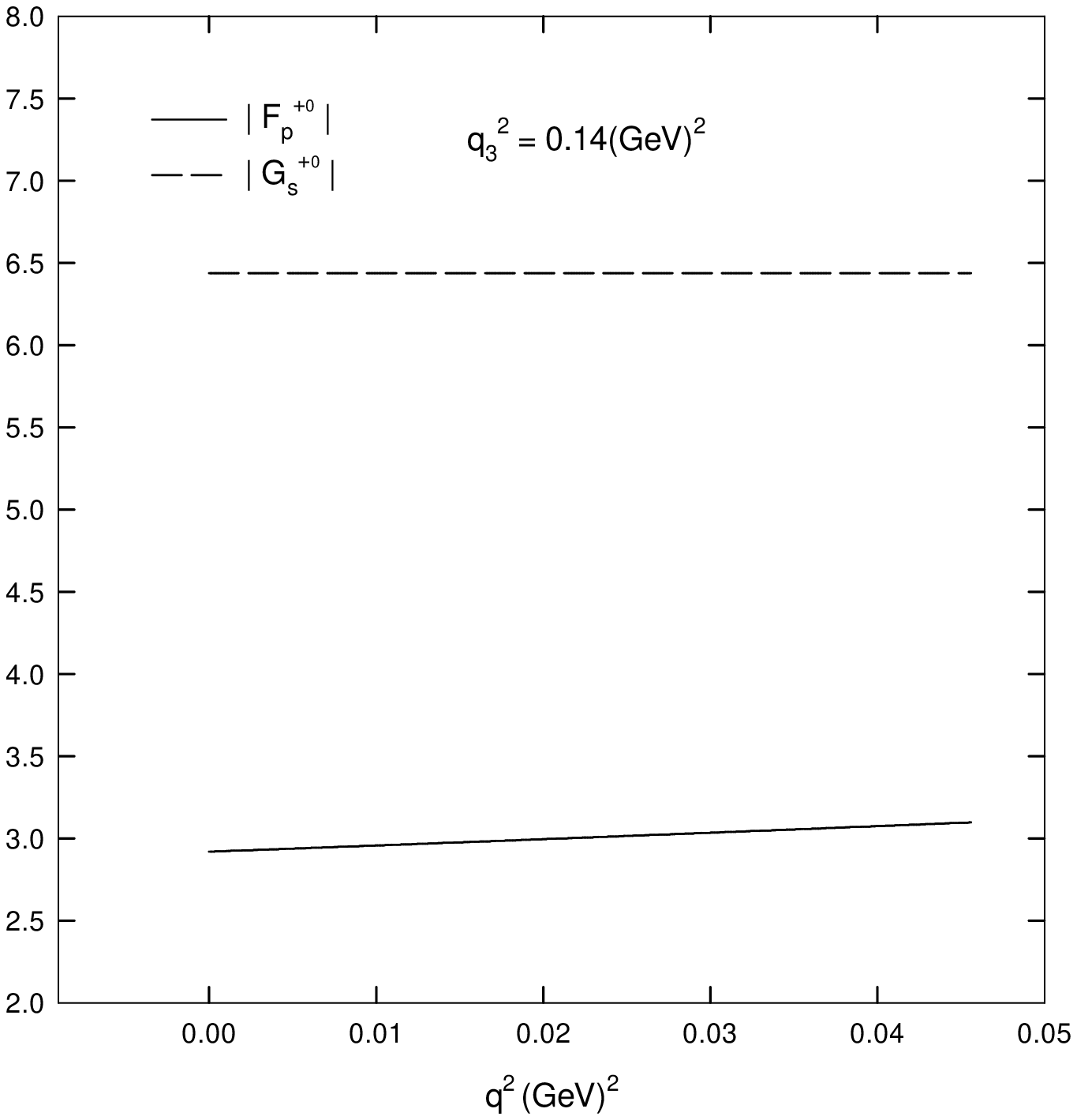}
Fig.9
\end{center}
\end{figure}
 
\begin{figure}
\begin{center}
\includegraphics[width=7in, height=7in]{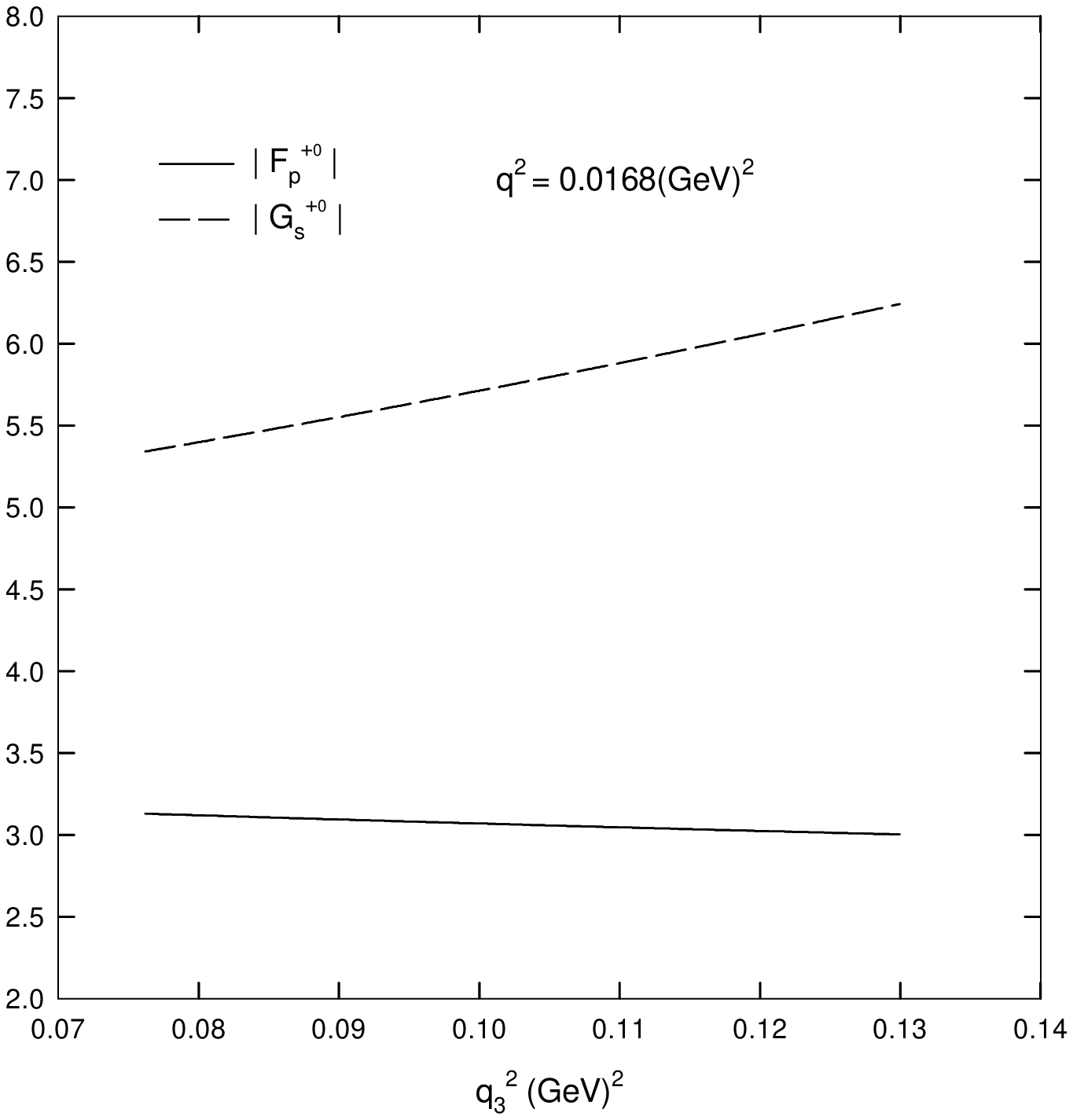}
Fig.10
\end{center}
\end{figure}

\begin{figure}
\begin{center}
\includegraphics[width=7in, height=7in]{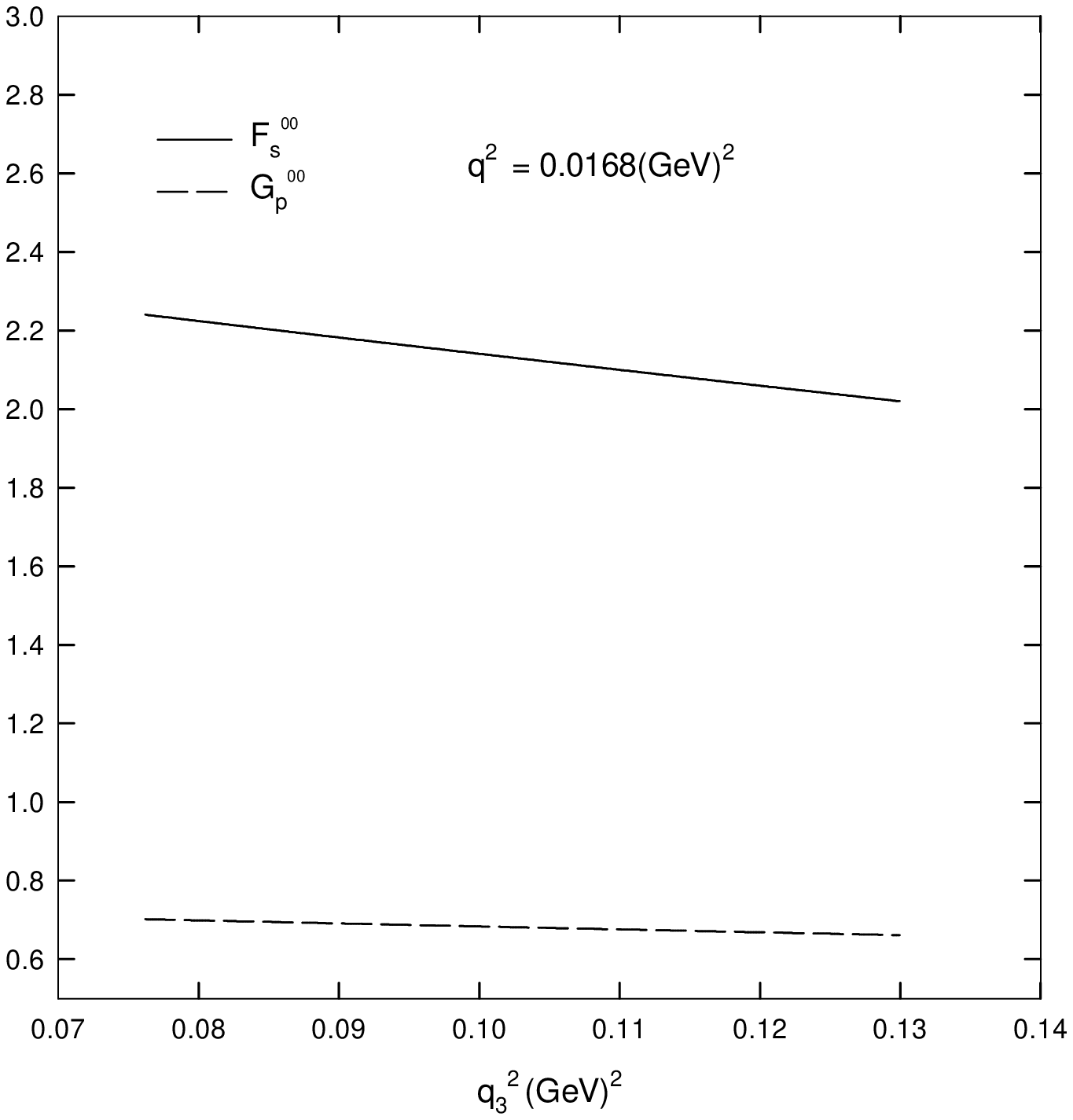}
Fig.11
\end{center}
\end{figure}
 
\begin{figure}
\begin{center}
\includegraphics[width=7in, height=7in]{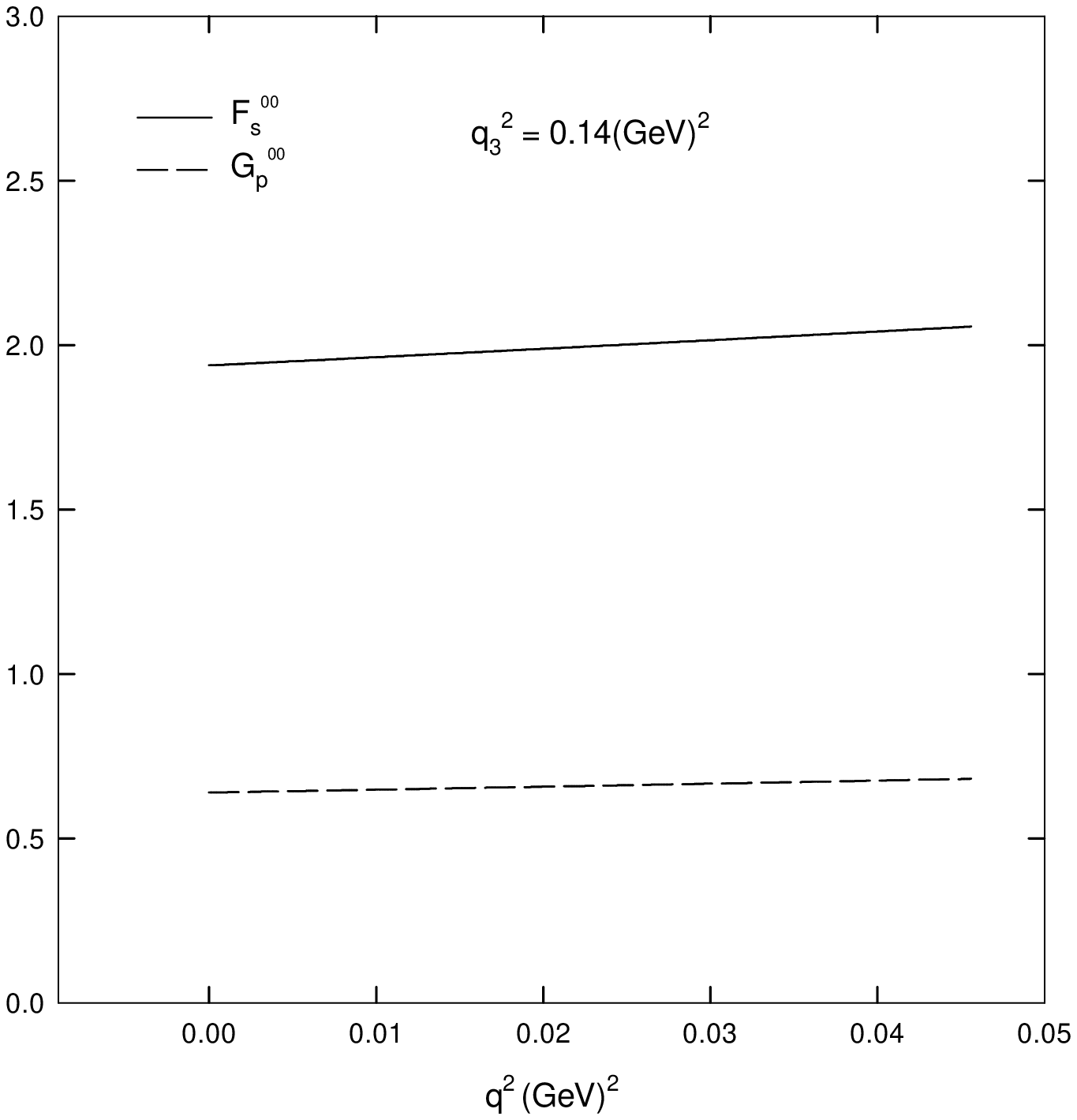}
Fig.12
\end{center}
\end{figure}


\begin{thebibliography}{40}
\bibitem{} A.Pais and S.B. Treiman, Phys. Rev.{\bf 168}, 168(1968);
L.M.Chounet, J.M.Gallard, and M.K.Gaillard, Phys. Rep. {\bf C4}, 199(1972).
Due to convention the form factors defined in Eq.(2) are the form factors
defined in[1] multiplied by ${1\over\sqrt{2}}$.
\bibitem{} N.Cabibbo and A. Maksymowicz, Phys.Rev. {\bf 137}, B 438(1965);
S. Weinberg Phys.Rev.Lett. {\bf 17}, 336(1966); J. Bijnens, Nucl. Phys.,
{\bf B337}, 635(1990); C. Riggenbach et al., Phys. Rev., {\bf D43}, 127(1991);
G.Colangelo, Phys.Lett., {\bf B336}, 543(1994).
\bibitem{} B.A.Li, Phys.Rev., {\bf D52}, 5165(1995), 5184(1995).
\bibitem{} G.'t Hooft, Nucl. Phys. {\bf B72}, 461(1974);{\bf B75}, 461(1974);
E.Witten, Nucl. Phys. {\bf B160}, 57(1979).
\bibitem{} B.A. Li, Phys. Rev. D {\bf 55}, 1436 (1997); {\bf 55} 1425 (1997).
\bibitem{} D.N. Gao, B.A. Li, and M.L. Yan, Phys. Rev. D {\bf 56}, 4115 (1997);
B.A. Li, D.N. Gao, and M.L. Yan, Phys. Rev. D {\bf 58}, 094031
(1998).
\bibitem{} J.Gao and Bing An Li, hep-ph/9911438.
\bibitem{} Bing An Li, hep-ph/9810311.
\bibitem{} Particle Data Group, Euro. Phys. J.{\bf C3},1(1998).
\bibitem{} L.Rosselet et al., Phys. Rev., {\bf D15}, 574(1977).
\bibitem{} V.V. Barmin et al., SJNP {\bf 55}, 547(1988).
\bibitem{} G.Makoff et al., Phys.Rev.Lett. {\bf 70},1591(1993).
\bibitem{} A.S.Carroll et al., Phys. Lett., {\bf 96B}, 407(1980).
\bibitem{} G.Doaldson, Thesis SLAC-0184.
 
 
 
 
 
 
\end{thebibliography}
\end{document}